\documentclass[twocolumn,aps,superscriptaddress,showpacs]{revtex4}

\usepackage{amssymb}
\usepackage{amsmath}
\usepackage{graphicx}
\usepackage[normalem]{ulem}
\usepackage[dvips]{color}

\begin{document}

\title{The effect of triangular flow on di-hadron azimuthal correlations in relativistic heavy ion collisions}
\author{Jun Xu}
\affiliation{Cyclotron Institute, Texas A\&M University, College
Station, TX 77843-3366, USA}
\author{Che Ming Ko}
\affiliation{Cyclotron Institute and Department of Physics and
Astronomy, Texas A\&M University, College Station, TX 77843-3366,
USA}

\date{\today}

\begin{abstract}
Using the AMPT model for relativistic heavy ion collisions, we have
studied the di-hadron azimuthal angular correlations triggered by
emitted jets in Au+Au collisions at center of mass energy
$\sqrt{s_{NN}}=200$~GeV and impact parameter $b=8$ fm. A double-peak
structure for the associated particles at the away side of trigger
particles is obtained after subtracting background correlations due
to the elliptic flow. Both the near-side peak and the away-side
double peaks in the azimuthal angular correlations are, however,
significantly suppressed (enhanced) in events of small (large)
triangular flow, which are present as a result of fluctuations in
the initial collision geometry. After subtraction of background
correlations due to the triangular flow, the away-side double peaks
change into a single peak with broad shoulders on both sides. The
away side of the di-hadron correlations becomes essentially a single
peak after further subtraction of higher-order flows.
\end{abstract}


\maketitle


Collisions of heavy nuclei at the Relativistic Heavy-Ion Collider
(RHIC) have created a hot and dense matter that is believed to
consist of deconfined quarks and gluons, as the inferred energy
density is much greater than the critical value of $\sim1$
GeV/fm$^3$ for the formation of a quark-gluon plasma
(QGP)~\cite{Ars05,Bac05,Ada05a,Adc05}. This matter was found to have
a very small viscosity and thus behaves like a nearly perfect
fluid~\cite{Gyu05,Shu05}. Among the signatures for the formation of
the QGP is the large quenching of energetic jets produced from
initial hard collisions, leading to the suppressed production of
hadrons with large transverse momenta~\cite{Adc01,Adl02}. Further
studies of the di-hadron azimuthal angular correlations triggered by
emitted jets have provided the possibility of investigating the
response of the QGP to the quenched away-side
jets~\cite{Adl03a,Adl03b,Ada04}. It was found in these studies that
for central and mid-central collisions of heavy nuclei, there
existed not only a pronounced peak at the near side of trigger jets
but also a broad double-peak structure at their away
side~\cite{Ada05b,Adl06}. Different mechanisms have been proposed to
explain this interesting phenomenon, including the medium-induced
gluon radiation~\cite{Vit05,Pol07}, the Mach cone from the shock
wave generated by the traversing jet in produced
QGP~\cite{Cas06,Ren06}, the path-length-dependent jet energy
loss~\cite{Chi06,Liw09}, the Cerenkov radiation from the
jet~\cite{Koc06}, the strong parton cascade~\cite{Mgl06}, and the
deflection of jets by partons in the QGP~\cite{Bet09,Bet10,Lih10}.
Also, it was recently pointed out that the so far overlooked large
triangular flow, resulting from the non-vanishing triangularity in
the initial collision geometry as a result of the spatial
fluctuation of participating nucleons~\cite{Pet10,Sch10,Qin10},
could be responsible for the double-peak structure at the away side
of di-hadron correlations~\cite{Alv10a,Alv10b}. This stems from the
observation that the azimuthal angular correlations of hadrons
resulting from their triangular flow have peaks at $0$, $2\pi/3$ and
$4\pi/3$, which are similar to the peaks observed in the
experimentally measured di-hadron azimuthal angular correlations. It
was shown that the away-side double peaks could indeed appear in the
hydrodynamic model if the initial density fluctuations are
included~\cite{Tak09}. However, a recent paper by the STAR
Collaboration~\cite{Aga10} shows that the away-side double-peak
structure is not much affected after subtracting the background
correlations that include the effect based on an estimated
triangular flow. It is therefore of great interest to use a
dynamical model to study consistently the effect of triangular flow
on di-hadron azimuthal angular correlations triggered by emitted
jets in relativistic heavy ion collisions. In this paper, we carry
out such a study within the framework of the AMPT
model~\cite{Zhang00,Lin05}.

The AMPT model is a hybrid model with the initial particle
distributions generated by the HIJING model~\cite{Wan91}. In the
version of string melting, which has been shown to better describe
the experimental observations at RHIC~\cite{Lin02a,Lin02b,Chen04},
hadrons produced from the HIJING model are converted to their
valence quarks and antiquarks, and their evolution in time and space
is then modeled by the ZPC parton cascade model~\cite{Zha98}. After
stopping scattering, quarks and antiquarks are converted via a
spatial coalescence model to hadrons, which are followed by hadronic
scattering via the ART model~\cite{Bal95}. As in previous studies~\cite{Lin02a,Lin02b,Chen04},
the elastic two-body scattering cross section used in the parton
cascade is set to be $10$ mb, which is larger than the value
expected from leading-order pQCD, to compensate for the neglected
higher-order inelastic processes.

\begin{figure}[h]
\centerline{\includegraphics[scale=0.8]{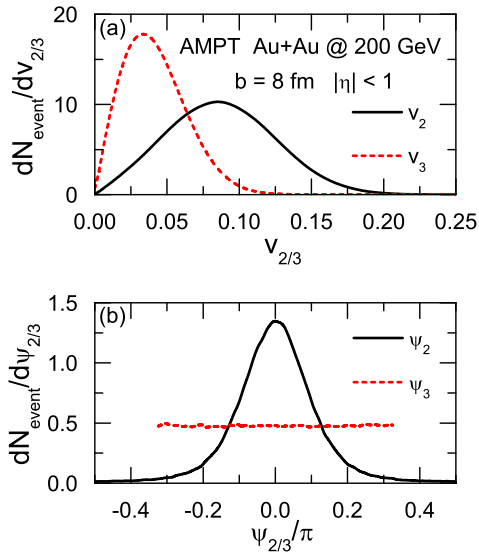}} \caption{(Color
online) Event distributions of elliptic flow $v_2$ and triangular
flow $v_3$ (panel (a)) and corresponding event plane angles $\psi_2$
and $\psi_3$ (panel (b)) in Au+Au collisions at $\sqrt{s_{NN}}=200$ GeV and $b=8$ fm from the
AMPT model. } \label{vpsi23}
\end{figure}

Using the AMPT model, we have studied Au+Au collisions at center of
mass energy $\sqrt{s_{NN}}=200$ GeV and impact parameter $b=8$ fm.
The latter is similar to the average value in minimum-bias
collisions that have been studied in experiments. The azimuthal
angle $\phi$ distribution of the $N(p_T)$ final hadrons with
transverse momentum $p_T$ in the transverse plane of a single AMPT
event can be written as~\cite{Alv10b}
\begin{equation}\label{fphi}
f(p_T,\phi) = \frac{N(p_T)}{2\pi} \{1+2
\sum_{n=1}^{+\infty} v_n(p_T)
\cos[n(\phi-\psi_{n}^{})]\},
\end{equation}
where the $n$th-order event plane angle $\psi_n$ and the anisotropic
flow $v_n(p_T)$ of hadrons with transverse momentum $p_T$ in this
event can be calculated, respectively, from
\begin{eqnarray}
\psi_n &=& \frac{1}{n} \arctan
\frac{\langle\sin(n\phi)\rangle}{\langle\cos(n\phi)\rangle},\\
v_n(p_T)
&=& \langle\cos[n(\phi-\psi_{n})]\rangle_{p_T},
\end{eqnarray}
with $\langle\cdots\rangle$ and $\langle\cdots\rangle_{p_T}$
denoting, respectively, average over all hadrons in the event and
over those of transverse momentum $p_T$. The total
$n$th-order anisotropic flow $v_n$ in this event is
then obtained by integrating $v_n(p_T)$ over transverse momentum.

In panel (a) of Fig.~\ref{vpsi23}, we show the event distributions
of the dominant elliptic flow $v_2$ and triangular flow $v_3$ of
hadrons in the mid-pseudorapidities $|\eta|<1$. It is seen that both
elliptic and triangular flows are large, although their values
change from event to event. The event distributions of the event
plane angles $\psi_2$ and $\psi_3$ of mid-pseudorapidity hadrons are
displayed in panel (b) of Fig.~\ref{vpsi23}. Similar to the results
from the ideal hydrodynamic model using initial fluctuating
collision geometry from the UrQMD model~\cite{Pet10}, the event
plane angle $\psi_2$ for the elliptic flow peaks at $0$ while the
event plane angle $\psi_3$ for the triangular flow is uniformly
distributed between $-\pi/3$ and $\pi/3$, as the former is mainly
due to the collision geometry while the latter is solely due to
fluctuations in the initial collision geometry. We have further
found that the distribution of $\psi_3$ remains similar for events
of fixed $\psi_2$, which shows that $\psi_3$ and $\psi_2$ are
essentially independent of each other.

\begin{figure}[h]
\centerline{\includegraphics[scale=0.8]{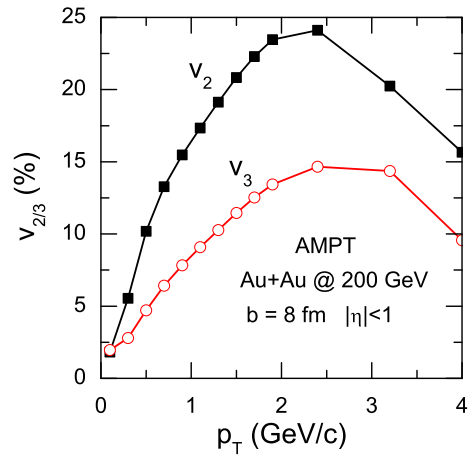}} \caption{(Color
online) Transverse momentum dependence of the hadron elliptic
($v_2$) and triangular ($v_3$) flows.} \label{vnpt}
\end{figure}

In Fig.~\ref{vnpt}, we show the transverse momentum dependence of
the event averaged elliptic and triangular flows of
mid-pseudorapidity hadrons, i.e., $\langle
N(p_T)v_n(p_T)\rangle_e/\langle N(p_T)\rangle_e$ with
$\langle\cdots\rangle_e$ denoting the average over all events, which
are the ones usually shown in the experimental results. It is seen
that both elliptic and triangular flows are appreciable at high
transverse momenta.

\begin{figure}[h]
\centerline{\includegraphics[scale=0.8]{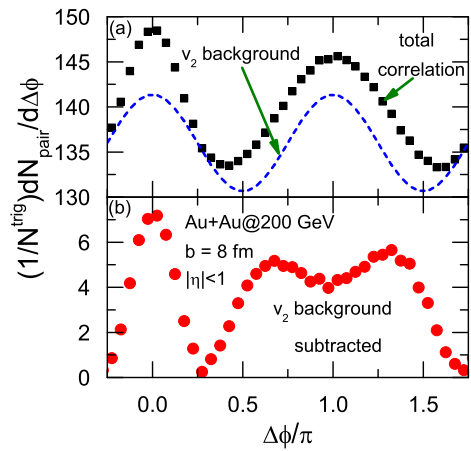}} \caption{(Color
online) Di-hadron azimuthal angular correlations per trigger
particle before (panel (a)) and after (panel (b)) subtracting
background correlations due to the hadron elliptic flow shown by the
dashed line in panel (a). } \label{corrtotal}
\end{figure}

The di-hadron azimuthal angular correlations between trigger
particles and associated particles are calculated from all such
pairs within the same event and then averaged over all events. With
trigger particles of transverse momenta in the window $2.5~{\rm
GeV}/c<p_T^{\rm trig}<6~{\rm GeV}/c$ and associated particles of
transverse momenta in the window $0.15~{\rm GeV}/c <p_T^{\rm
asso}<2.5~{\rm GeV}/c$, which are similar to the 'soft' case in
Ref.~\cite{Mgl06}, the result for the total di-hadron correlations
per trigger particle in the mid-pseudorapidity is shown by solid
squares in panel (a) of Fig.~\ref{corrtotal}. It is seen that the
associated particles have not only a near-side peak at
$\Delta\phi=0$ but also a broad away-side peak at $\Delta\phi=\pi$.

The di-hadron correlations include contributions from both the
back-to-back jet pairs and background anisotropic
flows~\cite{Pos98}. The latter can be obtained from the di-hadron
azimuthal angular correlations calculated from Eq.(\ref{fphi}),
i.e.,
\begin{eqnarray}\label{raw}
\left\langle\frac{dN_{\rm pair}}{d\Delta\phi}\right\rangle_e &=&
\frac{1}{2\pi}[\langle N^{\rm trig}N^{\rm asso}\rangle_e\notag\\
&+& 2 \sum_{n=1}^{+\infty} \langle N^{\rm trig}N^{\rm asso} v_n^{\rm
trig} v_n^{\rm asso}
\rangle_e \cos(n\Delta\phi)]\notag\\
\end{eqnarray}
by replacing $\langle N^{\rm trig}N^{\rm asso}\rangle_e$ with
$\langle N^{\rm trig}\rangle_e\langle N^{\rm asso}\rangle_e$ and
$\langle N^{\rm trig}v_n^{\rm trig}N^{\rm asso}v_n^{\rm
asso}\rangle_e$ with $\langle N^{\rm trig}v_n^{\rm
trig}\rangle_e\langle N^{\rm asso}v_n^{\rm asso}\rangle_e
\equiv\langle N^{\rm trig}\rangle_e v_n^{\rm trig}\langle N^{\rm
asso}\rangle_e v_n^{\rm asso}$, where $N^{\rm trig}$ and $N^{\rm
asso}$ are the number of trigger and associated hadrons,
respectively, and $v_n^{\rm trig}$ and $v_n^{\rm asso}$ are their
$n$th-order anisotropic flows averaged over all events, i.e.,
\begin{eqnarray}\label{back}
\frac{dN_{\rm pair}^{\rm back}}{d\Delta\phi}
&=&\frac{\langle N^{\rm trig}\rangle_e\langle N^{\rm asso}\rangle_e}{2\pi}\notag\\
&\times&[1+2\sum_{n=1}^{+\infty}v_n^{\rm trig}v_n^{\rm asso}
\cos(n\Delta\phi)].
\end{eqnarray}
We note that the di-hadron azimuthal correlations calculated from
Eq.(\ref{raw}) up to $n=5$ are already very close to the ones
calculated from hadron pairs shown in Fig.~\ref{corrtotal}(a).

Although the background correlations contain contributions from
anisotropic flows of all orders, only those due to the elliptic flow
have been considered in the experimental analysis. As shown by the
dashed line in panel (a) of Fig.~\ref{corrtotal}, the di-hadron
correlations due to the hadron elliptic flow show a similar peak
structure as in the total di-hadron correlations. The di-hadron
correlations after subtracting the background correlations due to
the elliptic flow are shown by solid circles in panel (b) of
Fig.~\ref{corrtotal}, and as in the experimental data they show a
double-peak structure at the away side of trigger particles.

\begin{figure}[h]
\centerline{\includegraphics[scale=0.8]{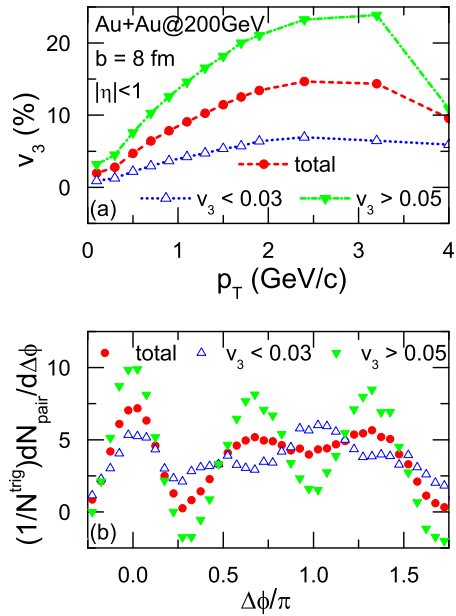}} \caption{(Color
online) (a) The triangular flow $v_3$ as a function of transverse
momentum and (b) di-hadron azimuthal
angular correlations per trigger particle from all events and from
events of small or large triangular flow.} \label{correps3abd}
\end{figure}

Before subtracting the background correlations due to the hadron
triangular flow, we consider separately events of small ($v_3<0.03$)
and large ($v_3>0.05$) triangular flows. In panel (a) of
Fig.~\ref{correps3abd}, the transverse momentum dependence of
triangular flows in these two cases are shown, respectively, by open
triangles and inverted triangles, and they are compared with that
for all events (solid circles). The di-hadron azimuthal angular
correlations after subtracting background correlations due to the
elliptic flow are shown in panel (b) of Fig.~\ref{correps3abd} for
these two cases. Compared with the result from all events, both the
near-side peak and the away-side double peaks in the di-hadron
correlations are significantly enhanced (suppressed) in events of
large (small) triangular flow, which shows that the triangular flow
has a large effect on the di-hadron correlations as suggested in
Refs.~\cite{Alv10a,Alv10b}.

We have also studied how the initial triangularity in the collision
geometry, which is defined as $\epsilon_3=\sqrt{\langle
r_{\text{init}}^3 \cos(3\phi_{\text{init}}) \rangle^2 + \langle
r_{\text{init}}^3 \sin(3\phi_{\text{init}}) \rangle^2}/ \langle
r_{\text{init}}^3 \rangle $~\cite{Pet10} with $r_{\text{init}}$ and
$\phi_{\text{init}}$ being the polar coordinates of initial partons,
affects the di-hadron azimuthal correlations. As in the above
consideration of the effect of the triangular flow, we find that
both the near-side peak and the away-side double peaks are
appreciably enhanced (suppressed) in events of large (small) initial
triangularity. This result thus indicates that in our model the
initial triangularity has a stronger effect on the di-hadron
azimuthal correlations than other physical mechanisms that can also
generate triangular flow.

\begin{figure}[h]
\centerline{\includegraphics[scale=0.8]{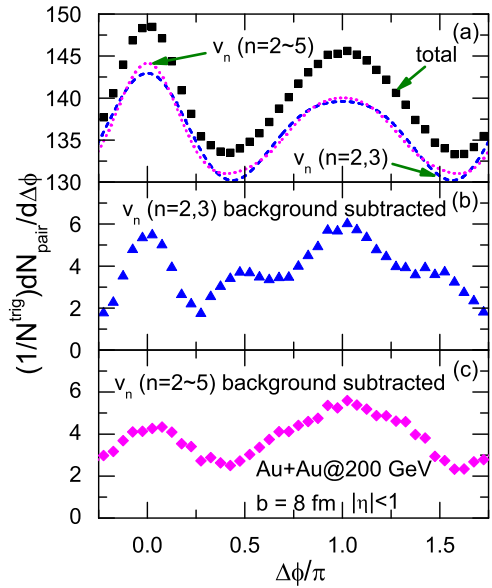}} \caption{(Color
online) Di-hadron azimuthal angular correlations per trigger
particle before (panel (a)) and after subtracting background
correlations due to both hadron elliptic and triangular flows (panel
(b)) shown by the dashed line in panel (a) and due to anisotropic
flows up to the 5th order (panel (c)) shown by the dotted line in
panel (a).} \label{subv23}
\end{figure}

The background correlations due to both hadron elliptic and
triangular flows are shown by the dashed line in panel (a) of
Fig.~\ref{subv23}. Subtracting these background correlations from
the total correlations results in the di-hadron correlations shown
in panel (b) of Fig.~\ref{subv23}. It is seen that the away-side
double peaks change into a single peak with broad shoulders on both
sides, and the near-side peak is also suppressed. The residual
correlations are similar to those from events of small triangular
flow as shown in panel (b) of Fig.~\ref{correps3abd}. After further
subtracting background correlations due to anisotropic flows up to
$n=5$ shown by the dotted line in panel (a), the away side of the
di-hadron correlations becomes essentially a single peak as shown in
panel (c) of Fig.~\ref{subv23}. The remaining near-side peak and the
away-side broad peak in the di-hadron azimuthal correlations are
then due to the effect of the back-to-back jet pairs produced in
initial hard scattering. Our results thus indicates that the effects
of elliptic and triangular flows on the di-hadron azimuthal
correlations are very important, while those of higher-order
anisotropic flows are not negligible either.

Although the away-side double-peak structure in di-hadron azimuthal
correlations is dominated by the triangular flow in heavy ion
collisions at $b=8$ fm, it was shown, however, in Ref.~\cite{Mgl10}
that the double-peak structure remains appreciable in central heavy
ion collisions at $b=0$ fm after subtracting background correlations
due to anisotropic flows even up to the 5th
order. This is likely due to the fact that the
anisotropic flows from initial density fluctuations are much weaker
in central collisions as a result of the larger particle
multiplicity and other effects such as those from jet deflections or
Mach cone shock waves are stronger.

In conclusion, we have investigated the di-hadron azimuthal angular
correlations triggered by emitted jets in the AMPT model with string
melting for Au+Au collisions at $\sqrt{s_{NN}}=200$ GeV with $b=8$
fm. Although the total di-hadron correlations show only a single
away-side peak besides a near-side peak, a double-peak structure
appears in the away side after subtracting the background
correlations due to the hadron elliptic flow. Both the near-side
peak and away-side double peaks in the di-hadron correlations are
found to be sensitive to the hadron triangular flow as they are
enhanced (suppressed) when only events of large (small) hadron
triangular flow are considered. Moreover, the away-side double peaks
change into a single peak after the subtraction of background
correlations due to both hadron elliptic and triangular flows.
Although other
effects~\cite{Vit05,Pol07,Cas06,Ren06,Chi06,Liw09,Koc06,Mgl06,Bet09,Bet10,Lih10}
on di-hadron correlations are appreciable in central collisions, the
away-side double peaks may be dominated by the triangular flow in
mid-central collisions.

\begin{acknowledgments}
This work was supported in part by the U.S. National Science
Foundation under Grant No. PHY-0758115 and the Welch Foundation
under Grant No. A-1358.
\end{acknowledgments}

\end{document}